\documentclass[aps,twocolumn,floatfix]{revtex4}

\usepackage{graphicx}
\usepackage{color}
\usepackage{times}

\newcommand{\Cnl}{C_{\rm nl}}
\newcommand{\D}{\mathbf{\mathcal{D}}}
\newcommand{\X}{\mathbf{X}}
\newcommand{\etal}{\emph{et al}}

\begin{document}

\title{Microcanonical temperature for a classical field: application to
Bose-Einstein condensation}
\author{M.~J.~Davis}
\email{mdavis@physics.uq.edu.au}
\affiliation{ARC Centre of Excellence for Quantum Atom Optics, 
Department of Physics, University of Queensland, St Lucia, QLD 
4072, Australia}
\author{S.~A.~Morgan}
\email{sam@theory.phys.ucl.ac.uk}
\affiliation{Department of Physics and Astronomy, University College London, 
Gower Street, London WC1E~6BT, United Kingdom}

\begin{abstract}

We show that the projected Gross-Pitaevskii equation (PGPE) can be mapped 
exactly onto Hamilton's equations of motion for classical position and momentum
variables.  Making use of this mapping, we adapt techniques developed in
statistical mechanics to calculate the temperature and chemical potential of a
classical Bose field in the microcanonical ensemble.  We apply the method to
simulations of the PGPE, which can be used to represent the highly occupied
modes of Bose condensed gases at finite temperature. The method is rigorous,
valid beyond the realms of perturbation theory, and agrees with an earlier
method of temperature measurement for the same system.  Using this method we
show that the critical temperature for condensation in a homogeneous Bose gas
on a lattice with a UV cutoff increases with the interaction strength. We
discuss how to determine the temperature shift for the Bose gas in the
continuum limit using this type of calculation, and obtain a result in
agreement with more sophisticated Monte Carlo simulations.  We also consider
the behaviour of the specific heat.

\end{abstract}

\maketitle

\section{Introduction} 

The Gross-Pitaevskii equation (GPE) has proven to be an extremely useful
description of macroscopic Bose-Einstein condensates (BECs) at or near zero
temperature \cite{dalfovo}.  It is the first and sometimes only tool to be used
in the description of many experiments in the field of non-linear atom optics
and Bose-Einstein condensation.  The validity of the GPE for many wide-ranging
experimental situations now appears beyond doubt.

However, it has been proposed that the GPE can also be used to represent the
non-equilibrium dynamics of Bose gases at finite temperature
\cite{boris,kagan1,kagan2,kagan3}.    The underlying argument is that for modes
of the gas with  an average occupation much larger than one, the classical
dynamics is far more important than the quantum dynamics.  This is analogous to
the semi-classical approximation utilised in laser physics for the
electromagnetic field.   A major advantage of using the GPE in such
situations is that it is non-perturbative, and so can be applied in the region
of the critical point as long as the condition on occupation numbers is
observed.
In Ref.~\cite{davis2001a} a finite temperature
Gross-Pitaevskii equation is derived from the quantum many-body Hamiltonian for
the Bose gas with this approximation in mind.  An alternative route to similar
equations of motion is possible via the use of the Wigner representation
\cite{steel1998}.  This
approach may be more familiar to those from the quantum optics community.

Some of the first numerical
calculations utilising the GPE for finite temperature simulations
were performed by Damle \etal.\ \cite{damle}
and Marshall \etal.\ \cite{Marshall}.
More recently there have been several calculations using the so-called 
``classical field''
approximation. In particular we mention those of the Warsaw group
\cite{goral1,goral2,schmidt}, the ENS group
\cite{sinatra1,sinatra2,sinatra3,lobo}, and the current authors 
\cite{davis2001b,davis2002}.  These calculations are all related by
the fact that they include no damping terms in the GPE, and thus rely on
\emph{ergodicity} for the system to thermalize.
 Classical field methods involving
both damping and stochastic terms have been considered by Gardiner \etal.\
\cite{stochasticGPE}, Stoof and Bijlsma \cite{bijlsma}, and
 Duine and Stoof \cite{duine2001a}.

  While the qualitative results from previous work have been promising, there
has been some difficulty in performing quantitative calculations using such
methods, and in particular in determining the temperature of the system at
equilibrium. We partially addressed this issue in previous work using a variety
of methods  to determine the temperature of our simulations
\cite{davis2001b,davis2002}. The most reliable of these involved fitting
time-averaged quasiparticle occupations and energies to the classical limit of
the Bose-Einstein distribution function.  However, this method relies on the
existence of a basis which approximately  diagonalizes the Hamiltonian
(quasiparticles) for which energies and wave functions  can be calculated in
advance. This method is therefore only applicable in the realm of perturbation
theory, and fails for even moderate temperatures in systems with large
nonlinearities.  Hence it is desirable to find a more widely applicable scheme
for unambiguously determining the temperature of numerical simulations.

 In a succinct yet insightful paper, Rugh \cite{rugh97} expressed the 
temperature of a classical Hamiltonian system in terms of a phase space 
expectation value of a suitable function of the canonical position and momentum
coordinates [Eq.~(\ref{eqn:Rugh_T}) of this paper.] Using the ergodic
theorem, this expectation value over phase space can be interpreted as a
dynamical average for a system in equilibrium, and immediately lends itself to
application in numerical calculations.  Rugh developed this procedure further
in \cite{rugh98}, and generalised it to include systems with other conserved
quantities in addition to the energy \cite{rugh01}. This generalization turns
out to be crucial for the  application of the method to the interacting
Bose gas.  Rugh's formula for the temperature has been applied to several
systems to date, for example in the field of molecular dynamics.  This
has led to the notion of a configurational temperature for gases, which only
depends on the spatial co-ordinates of the particles, in addition to the
usual kinetic temperature which only depends on the momenta
\cite{butler1998,jepps2000,delhommelle2001}.

In this paper we apply the  microcanonical formalism of Rugh to the BEC
Hamiltonian to determine the temperature of numerical simulations of thermal
Bose gases. The method is non-perturbative and does not rely on the existence
of well-defined Bogoliubov quasiparticles. The paper is organised as follows. 
In Sec.~\ref{sec:formula} we briefly summarise the expression for the
temperature and other derivatives of the microcanonical entropy, and describe their
application to the BEC Hamiltonian.  Section \ref{sec:numerical} presents our
numerical results for our projected GPE (PGPE) system, while
Sec.~\ref{sec:other} relates our calculations to other dynamical calculations
of classical  $\phi^4$ field theory, as well as to calculations of the shift of
the transition temperature for a homogeneous Bose gas.  We conclude in
Sec.~\ref{sec:conc}.

\section{Formalism}
\label{sec:formula}

\subsection{Hamiltonian}

We consider a classical system with $M$ independent modes.  The  Hamiltonian can
be written as $H = H(\mathbf{\Gamma})$, where $\mathbf{\Gamma} = \{\Gamma_i\} =
\{Q_i,P_i\}$ is a vector of length $2M$ consisting of the canonical position
and momentum co-ordinates.  In these coordinates we define the gradient
operator $\nabla$ in terms of its components $\nabla_i = \partial/\partial\Gamma_i$.

In the notation of Rugh \cite{rugh01}, the Hamiltonian $H$ may have a number of
independent first integrals, labelled by $F = F_1, \ldots,F_m,$ 
that are invariant under the dynamics of $H$.  We could define
$F_0 = E$, and include the conserved energy with the other
constants of motion in this notation, but for clarity we consider it separately. A particular macro-state of
such a system can be specified by the values of the conserved quantities,
labelled as $H = E, F_i = I_i$.

The expression for the temperature of 
such a system in the microcanonical ensemble is given by
\begin{equation}
\frac{1}{k_B T} 
= 
\left(\frac{\partial S}{\partial E}\right)_{F_i},
\end{equation}
where all other constants of motion are held fixed, and where the entropy is
given by
\begin{equation}
e^{S/k_B} =
 \int d\mathbf{\Gamma} \;\delta[E - H(\mathbf{\Gamma})]
\; \prod_i \delta[{I_i -
 F_i(\mathbf{\Gamma})}].
\end{equation}

In this case,  the temperature of the
system can be written as
\begin{equation}
\frac{1}{k_B T} 
= 
\bigg\langle \mathbf{\D} \cdot \mathbf{X}(\mathbf{\Gamma}) \bigg\rangle,
\label{eqn:temp_eqn}
\end{equation}
where the angle brackets correspond to an ensemble average, and the components of the vector operator $\D$ are
\begin{equation}
\D_i = e_i \frac{\partial}{\partial \Gamma_i},
\end{equation}
where $e_i$ can be chosen to be any scalar value, including zero.
The vector field $\mathbf{X}$ can also be chosen freely within the constraints
\begin{eqnarray}
\D H \cdot \mathbf{X} =1,
\quad \D F_i \cdot \mathbf{X} = 0, \quad 1\le i \le m.
\label{eqn:X_conditions}
\end{eqnarray}
Geometrically this means that the vector field $\mathbf{X}$ has a non-zero
component transverse to the $H = E$ energy surface, and is parallel to the surfaces
$F_i = I_i$. The expectation value in  
Eq.~(\ref{eqn:temp_eqn}) is over all possible states in the microcanonical ensemble;
however if the ergodic theorem is applicable then it can equally well be interpreted
as a time-average.  For further details on the origin of this expression we refer
the reader to  Rugh's original papers \cite{rugh97,rugh98,rugh01}, as well as
derivations  found in Giardin\`{a} and Levi \cite{giardina1998}, Jepps \etal.\
\cite{jepps2000} and Rickayzen and Powles \cite{rickayzen2001}.

\subsection{Dimensionless BEC Hamiltonian}
The full quantum many-body Hamiltonian for the Bose gas in dimensionless form is 
\begin{eqnarray}
\tilde{H} = \int d^3 \tilde{\mathbf{x}} \left[  \nabla
\tilde{\Psi}^{\dag}(\tilde{\mathbf{x}}) \cdot \nabla \tilde{\Psi}(\tilde{\mathbf{x}}) +
\tilde{V}(\tilde{\mathbf{x}})\tilde{\Psi}^{\dag}(\tilde{\mathbf{x}})\tilde{\Psi}(\tilde{\mathbf{x}}) +
\right.\nonumber\\
\left.\frac{\Cnl}{2}
\tilde{\Psi}^{\dag}(\tilde{\mathbf{x}})\tilde{\Psi}^{\dag}(\tilde{\mathbf{x}})
\tilde{\Psi}(\tilde{\mathbf{x}})\tilde{\Psi}(\tilde{\mathbf{x}})\right],
\label{eqn:qH}
\end{eqnarray}
where $H = N \epsilon_L \tilde{H}$, $N$ is the number of particles in the system, $\tilde{\mathbf{x}} = 
\mathbf{x}/L$, 
$L$ is the unit of length, 
$\epsilon_L = \hbar^2/(2mL^2)$ is the unit of energy, $m$ is the mass of the
particles and 
$\tilde{V}(\tilde{\mathbf{x}})$ is the
dimensionless external potential if any is present.
The dimensionless quantum Bose field operator $\tilde{\Psi}(\tilde{\mathbf{x}})$ 
is here normalized to one,
 $\int d^3 
  \tilde{\mathbf{x}} \langle \tilde{\Psi}^{\dag}(\tilde{\mathbf{x}}) 
  \tilde{\Psi}(\tilde{\mathbf{x}}) \rangle=1$,
and $\Cnl$ is the nonlinear constant defined as
\begin{equation}
\Cnl = \frac{N U_0}{\epsilon_L L^3} = \frac{8\pi a N}{L},
\end{equation}
where $a$ is the \emph{s}-wave scattering length.  In this expression we have
assumed a high momentum cutoff and  made use of the replacement of the true
interatomic potential with the two body T-matrix, $V(\mathbf{x}) \rightarrow
U_0 \delta(\mathbf{x})$, where $U_0 = 4\pi \hbar^2 a / m $.

In Ref.~\cite{davis2001a} the field operator is split into a classical part and a
quantum part, with the boundary determined by the requirement that the average
occupation number 
$\langle N_k \rangle$ of modes below the cutoff satisfies
$\langle N_k \rangle \gg 1$.
Equations of motion were derived for the classical part, before taking the mean value. 
This resulted in the finite temperature Gross-Pitaevskii equation (FTGPE),
which describes the evolution of a classical field coupled to an effective bath
described by a quantum Boltzmann-like equation.  This equation proves to be
somewhat difficult to solve numerically, and in
Refs.~\cite{davis2001b,davis2002} we  reported results focussing on a
simplification we termed the projected Gross-Pitaevskii equation (PGPE).  This
equation describes the evolution of a  classical field only, with a
cutoff at a given momenta or energy. 
It is identical to the usual GPE except that it evolves a wave function
which is restricted to a finite-sized basis satisfying the classical condition
$\langle N_k \rangle \gg 1$.  The PGPE for a homogeneous system is written
explicitly for the homogenous gas in Eq.~(\ref{eqn:pgpe}) of this paper.

In this paper we wish to determine the temperature of the restricted system
described by the evolution of the PGPE.  Thus the Hamiltonian we consider is the
classical version of Eq.~(\ref{eqn:qH}) obtained by replacing the field operator 
 $\tilde{\Psi}(\mathbf{\tilde{x}})$ with the classical field 
 $\psi(\tilde{\mathbf{x}})$,
subject to the important restriction that $\psi(\tilde{\mathbf{x}})$
 is constructed from a 
finite number of low-energy modes. We can therefore write it in the form
\begin{equation}
\psi(\tilde{\mathbf{x}}) = \sum_{k \in C} c_k \phi_k(\tilde{\mathbf{x}}),
\end{equation}
where $C$ labels the classical modes in the coherent region below the
cutoff, as defined in
\cite{davis2001a}.

\subsection{Canonically conjugate position and momentum variables}

We must now make a choice of the canonically conjugate co-ordinates of our
Hamiltonian.  As we are defining our classical field in a basis, 
it seems natural to convert to a basis representation.  If we choose our basis to be
that which diagonalises the ideal gas Hamiltonian 
[the first two terms of Eq.~(\ref{eqn:qH})]
we find
\begin{equation}
H = \sum_n \epsilon_n c_n^* c_n
+ \frac{\Cnl}{2} \sum_{mnpq} 
c_m^* c_n^* c_p c_q \langle m n|p q\rangle,
\label{eqn:Hbasis}
\end{equation}
where the matrix element is
\begin{equation}
\langle m n | p q \rangle = \int d^3\mathbf{x} \,
\phi^*_m(\mathbf{x}) \phi^*_n(\mathbf{x})   \phi_p(\mathbf{x})  \phi_q(\mathbf{x}).
\label{eqn:matrix_element}
\end{equation}
The equation of motion for the $\{c_n \}$ is given by the PGPE.  This problem
can be mapped exactly to the one considered by Rugh by defining real,
canonically-conjugate coordinates $Q_n$ and $P_n$ 
\begin{equation}
Q_n = \frac{1}{\sqrt{2 \epsilon_n}}(c_n^* + c_n), \quad
P_n = i\sqrt{\frac{\epsilon_n}{2}}(c_n^* - c_n),
\label{def:XP}
\end{equation}
with the corresponding inverse transformation
\begin{equation}
c_n = \sqrt{\frac{\epsilon_n}{2}} Q_n + \frac{i}{\sqrt{2 \epsilon_n}} P_n, \quad
c_n^* = \sqrt{\frac{\epsilon_n}{2}} Q_n - \frac{i}{\sqrt{2 \epsilon_n}} P_n.
\end{equation}
With these definitions, the evolution of the $c_n$ coefficients given by the
PGPE maps exactly to the evolution of the coordinates $Q_n$ and $P_n$ given by
Hamilton's equations. The PGPE is therefore in one-to-one correspondance with a
classical microcanonical system and its equilibrium propreties can be studied
using the wide variety of techniques which have been developed in classical
statistical mechanics.

We have performed numerical calculations for the homogeneous PGPE, and so we use a plane wave
basis  where $\phi_n(\tilde{\mathbf{x}})$ = $\exp(i \tilde{\mathbf{k}}_n\cdot
\tilde{\mathbf{x}})$,  $n = \{n_x,n_y,n_z\}$ and $\epsilon_n = |\tilde{k}_n|^2 =
(2\pi |n|)^2$. 
However, the method we describe is general and can be applied
directly to inhomogeneous systems for BECs in magnetic and optical dipole
traps.

\subsubsection*{Calculations on a grid}

The implementation of a projection operator in the GPE is an essential feature
of any classical simulation. While we have explicitly defined a
projection operator in terms of a basis set, other authors have implicitly
chosen a momentum cutoff by the use of a finite-size grid in their GPE
simulations \cite{goral2,schmidt}.  The method of temperature determination
described in this paper can also be applied to these calculations, but with a
different choice of postion and momentum co-ordinates.

On a finite grid, the Hamiltonian (\ref{eqn:qH}) can be discretised in real
space and the classical equivalent can be written as
\begin{eqnarray}
H= h_x h_y h_z\sum_{n}\bigg[ (\nabla \alpha_{n})^2 + (\nabla \beta_{n})^2 + 
 \nonumber\\
 V_{n}(\alpha_{n}^2 +
\beta_{n}^2) + \frac{C_{nl}}{2}(\alpha_{n}^2+\beta_{n}^2)^2 \bigg],
\label{eqn:Hgrid}
\end{eqnarray}
where $n\equiv \{n_x,n_y,n_z\}$ labels the grid point, $h_x, h_y, h_z$ are the
grid spacings for each axis, and we have defined
$\psi_n = \alpha_n + i \beta_n$.  In this case the appropriate position and
momentum variables are
\begin{equation}
Q_n = \sqrt{2}\alpha_n, \qquad
P_n = \sqrt{2}\beta_n.
\end{equation}

However, we believe that it is important to define the projector using a basis
that is relatively well-defined in energy \emph{at the cutoff}  (we stress that this does not
mean that the basis has to be well-defined in energy \emph{below} the cutoff).
It has previously been shown \cite{crispin} that the single particle energy
levels of a partially condensed system are essentially those of the trapping
potential for energies $\epsilon \ge E_R \approx 3 \mu_C$, where $\mu_C$ is the condensate
energy eigenvalue.  Thus the above cutoff projector can be written
\begin{eqnarray}
\hat{\mathcal{Q}}\{F(\mathbf{x}) \}
&=& \sum_{k \notin C}\phi_k({\bf x}) \int d^3{\bf x}' \;\phi_k^*({\bf x}')
F({\bf x}'),
\label{eqn:projectorQ}
\end{eqnarray}
where the $\{\phi_k\}$ are the basis states appropriate to the potential, and the
notation $k \notin C$ describes a summation over all modes above the energy
cutoff $E_R$.  As this basis is
complete, the below cutoff projector is simply
\begin{equation}
\hat{\mathcal{P}} = \hat{1} - \hat{\mathcal{Q}},
\label{eqn:projectorP}
\end{equation}
which gives the result written explicitly in Eq.~(\ref{eqn:projector}).  
We also require the classical condition
\begin{equation}
N_k = \frac{k_B T}{E_R - \mu} \gg 1,
\end{equation}
to hold at the cutoff and so for $E_R \approx 3 \mu_C$ this should also be satisfied.

For a trapped Bose gas, the implicit momentum projector based on the
finite-grid method is not at all well-defined in energy at the cutoff, and we
believe that this may lead to difficulties.   However, this is yet to be
invesigated numerically; for further discussion of this issue we refer the
reader to \cite{Gardiner2003}.

\subsection{Choice of vector field $\mathbf{X}$}
\label{blergh}
In order to satisfy the conditions (\ref{eqn:X_conditions}) we can 
choose a vector field of the form
\begin{equation}
\mathbf{X} = a \D H + \sum_{i=1}^{m} b_i  \D F_i,
\label{eqn:X_expansion}
\end{equation}
where the $m+1$ coefficients $\{a,b_i\}$
are determined by the $m+1$ simultaneous equations in Eq.~(\ref{eqn:X_conditions}). 
Due to the freedom in the choice of 
the vector operator $\D$
 we can set any component of the length $2M$ vector $\mathbf{X}$ to zero.
This
turns out to be useful as the components corresponding to the momentum and
position variables can be different orders of magnitude.  
Two particular choices
we make use of later are $\X_P$ with $\D = \D_P = \{0,\partial/\partial P_i\}$
and $\X_Q$
with $\D = \D_Q= \{ \partial/\partial Q_i,\,0\}.$
These lead to two different calculations for the temperature that only agree
in general if the system is in thermal equilibrium.  This provides a useful check
that the simulations have in fact thermalized.

In Rugh's first two papers \cite{rugh97,rugh98} the only first integral
considered was the energy, and he chose $\D \equiv \nabla$ which yielded
the (dimensionless) formula
\begin{equation}
\frac{1}{T} = \bigg\langle \nabla \cdot \frac{\nabla H}{|\nabla H|^2}
\bigg\rangle. 
\label{eqn:Rugh_T}
\end{equation}
For the BEC Hamiltonian we consider, however, there are other first 
integrals that must be
taken into account.  Most importantly, 
the evolution conserves the normalization of the wave function,
but other first integrals that may occur are both the angular and linear momentum.

The effect of including these additional first integrals in the definition of
the vector field $\mathbf{X}$ is to account for the energy that is
associated with a conserved quantity and hence is unavailable for
thermalization. This ensures that only the appropriate free energy is used to
calculate the temperature. We conjecture that the same result can be achieved
by first transforming to a co-ordinate system  where the total angular and
linear momenta, etc, are all zero and therefore do not contribute to the energy
of the system. In fact, Rugh demonstrated this explicitly in \cite{rugh01} for
a system of particles with a conserved centre-of-mass motion.


An exception to this, however, is the conservation of normalization 
$N = \sum_n c^*_n c_n$. This must be considered explicitly because
 there is no co-ordinate system in which it can be made to vanish. 
 The constraint on $N$ means that
the ground state of the system will, in general, have a finite energy. For
example, a non-interacting gas in a harmonic trap of frequency $\omega$ must
have at least the zero-point energy $\hbar \omega/2$ for each spatial degree of
freedom. For a non-ideal, homogeneous gas the restriction that at least one of
the $c_n$ must be non-zero means that there will always be a finite
interaction energy associated with the ground state energy $\tilde{E}_0
= \Cnl/2$. These energy contributions
are not accessible for thermalization, however, and including the normalization
constraint allows them to be removed. We note, however, that the effect of this
constraint is in general more complicated than a simple subtraction of the
ground state energy (which could be achieved by hand) and depends on the
definition of the operator $\D$ used to calculate the temperature, as shown
below.

To deal with the normalization constraint, we need to choose a vector field
$\mathbf{X}$ which satisfies Eqs.~(\ref{eqn:X_conditions}) with $F_1 = N = \sum_n
c^*_n c_n$. The result is
\begin{equation}
\mathbf{X} = \frac{\D H - \lambda_N\D N}
{|\D H|^2 - \lambda_N\D N \cdot \D H},
\label{eqn:Xvec}
\end{equation}
where the parameter
\begin{equation}
\lambda_N = \frac{\D N \cdot \D H}{|\D N|^2},
\label{eqn:X_with_N}
\end{equation}
looks similar to a chemical potential.  For a system on a real space grid with
$\D = \nabla$ and a Hamiltonian given by Eq.~(\ref{eqn:Hgrid})   we find that
$\lambda_N = \mu_{\rm GPE}$, where $\mu_{\rm GPE}$ is the usual Gross-Pitaevskii
form  of the chemical potential, obtained from the Hamiltonian of 
Eq.~(\ref{eqn:Hgrid}) by doubling the interaction term.  However, in general the expression of Eq.~(\ref{eqn:X_with_N}) 
does not have a simple interpretation.

Substituting Eq.~(\ref{eqn:Xvec}) into Eq.~(\ref{eqn:temp_eqn}) we find that 
our full expression for the temperature is
\begin{widetext}
\begin{eqnarray}
\frac{1}{T} = \left\langle
\frac{\D^2 H - \lambda_N \D^2N - \D\lambda_N\cdot\D N}
{|\D H|^2 - \lambda_N (\D H \cdot \D N)}
\right\rangle
-
\left\langle
\frac{ (\D H - \lambda_N\D N) \cdot [ \D |\D H|^2 - (\D H \cdot \D N) \D \lambda_N
 - \lambda_N \D (\D H \cdot \D N)]}
{[{|\D H|^2 - \lambda_N (\D H \cdot \D N)}]^2}
\right\rangle.
\label{eqn:T}
\end{eqnarray}
\end{widetext}
The second term in this expression is of order $1/M$, and so
in many situations it can reasonably be neglected.  However we have calculated
the full expression 
for all results presented in this paper.

\subsection{Other thermodynamic quantities}

The method described in this paper can also be used to calculate first derivatives of
the microcanonical entropy with respect to any of the first integrals of the
Hamiltonian \cite{rugh01}. In particular, we find to calculate the quantity
$$\left(\frac{\partial S}{\partial F_j}\right)_{E,F_i},\quad i\neq j,
$$the constraints on our vector field should be
\begin{eqnarray}
\D H \cdot \mathbf{X} = 0,
\quad \D F_i \cdot \mathbf{X} = 1, \quad \D F_j \cdot \mathbf{X} = 0, \quad i\neq j.
\label{eqn:X_conditions2}
\end{eqnarray}
For the BEC Hamiltonian, we have
\begin{equation}
\frac{\mu}{k_B T} = -\left(\frac{\partial S}{\partial N}\right)_{E},
\end{equation}
and implementing the required constraints, we find an appropriate vector
field is that given by Eq.~(\ref{eqn:Xvec},\ref{eqn:X_with_N}) but with the
roles of $H$ and $N$ reversed.

In addition, higher order derivatives of the
entropy can also be determined, making available quantities such as the
specific heat $c_{\rm sp}$ of the system \cite{rugh98}. This quantity could in principle be
calculated from the expression
\begin{equation}
\frac{1}{c_{\rm sp}} = 1 - \frac{\langle \nabla \cdot ( \X \nabla \cdot \X) \rangle}
{\langle\nabla \cdot \X\rangle^2},
\end{equation}
where the vector $\X$ is
 determined by Eq.~(\ref{eqn:Xvec},\ref{eqn:X_with_N}).
However, for the BEC Hamiltonian the expressions for such
quantities are unreasonably complicated, and we do not consider them in this paper. 
Instead, higher derivatives will simply be obtained numerically once the  temperature
is determined.

\section{Numerical results}
\label{sec:numerical}

In this section we apply the new formula Eq.~(\ref{eqn:T}) to data from  simulations of the
PGPE described in \cite{davis2001b,davis2002}, as well as to many new simulations
with a wider range of energies and nonlinear parameters $\Cnl$.
For a full description we refer the reader to
Ref.~\cite{davis2002}.  Briefly, the calculations evolve the projected Gross-Pitaevskii
equation \cite{davis2001a} for the homogeneous gas in three dimensions
\begin{equation}
i \frac{\partial \psi(\tilde{ \bf \mathbf{x}})}{\partial \tau}
= -\tilde{\nabla}^2 \psi( \tilde{\bf \mathbf{x}}) + 
C_{\rm nl} \hat{\mathcal{P}} \{|\psi(\tilde{ \bf \mathbf{x}})|^2\psi(\tilde{ \bf \mathbf{x}})\}.
\label{eqn:pgpe}
\end{equation}
The nonlinear constant is $C_{\rm nl} = {2 m N U_0}/{\hbar^2 L}$, where $N$ is
the total number of particles in the volume, and $L$ is the period of the
system.  Our dimensionless parameters are $\tilde{\bf \mathbf{x}} = {\bf \mathbf{x}}/L$, wave
vector  $\tilde{\bf k} = {\bf k} L$, energy  $\tilde{\varepsilon} = \varepsilon
/ \varepsilon_L$, and  time $\tau = \varepsilon_L t/ \hbar$, with 
$\varepsilon_{L} = \hbar^2/(2 m L^2)$. 
The projection operator 
$\hat{\mathcal{P}}$ excludes all components of the nonlinear term in the GPE
outside the coherent region, and is defined by 
[c.f. Eqs.~(\ref{eqn:projectorQ},\ref{eqn:projectorP})]
\begin{eqnarray}
\hat{\mathcal{P}}\{F(\mathbf{x}) \}
&=& \sum_{k \in C}\phi_k({\bf x}) \int d^3{\bf x}' \;\phi_k^*({\bf x}')
F({\bf x}'),
\label{eqn:projector}
\end{eqnarray}
where $\{\phi_k\}$ is an orthonormal basis appropriate to the problem. 
For the homogeneous system with periodic boundary conditions the
relevant basis is the plane wave states, and so  this procedure is simply the
application of a forward fourier transform, removal of components with $\tilde{k} >
\tilde{k}_c$, followed by the inverse transformation.  The quantity $\tilde{k}_c$ defines the
momentum cutoff
for the coherent region, and for all data presented in this paper we use $\tilde{k}_c = 15
\times 2 \pi$.

 We begin with randomised initial fields
$\psi( \tilde{\bf \mathbf{x}})$ with a given 
energy on a 3D grid with 32 points
in each dimension, and evolve these until the
field has reached equilibrium.  We calculate all thermodynamic quantities from sampling two
hundred field configurations in equilibrium. 

\subsubsection*{Cutoff dependence of simulations}
The choice of momentum cutoff $\tilde{k}_c = 15 \times 2 \pi$ is motivated
simply by computational convenience.  It also allows for comparison of the Rugh
method of temperature measurement with data from earlier calculations.

For a given initial energy, the resulting equilibrium temperature depends
on the number of modes below the cutoff.  This can be easily understood from
the equipartition theorem --- if more modes are present, less energy will be
contained in each one and therefore the final temperature will be lower. 
Also, the dimensionless 
critical temperature for a system with a fixed normalisation
depends on the cutoff, as can be seen in the text beneath Eq.~(\ref{eqn28}).

Work is currently in progress to develop a description of the modes above
the  cutoff and their coupling to the PGPE. The aim of this work is a
complete computational method which will be insensitive to the exact
position of the cutoff. Exploring and developing techniques for the
non-perturbative classical field is an important part of this programme,
and we focus on this aspect of the problem in this paper.

Despite this, there are some equilibrium calculations which can be
carried out immediately using an approximate treatment of the modes above
the cutoff. We present results for one such calculation (the shift in $T_c$
with interaction strength) in this paper. These results have only a weak
dependence on the cutoff.

\subsubsection*{Use of the classical field method at $T_c$}

The classical field can only describe modes that satisfy the high
occupation condition.  But even at the critical temperature and above, the
lowest energy states will have the largest occupations --- and for a wide range
of parameters, many  of these can satisfy $N_k \gg 1$. These are the modes that
are responsible for critical behaviour,  such as the shift in $T_c$ and the
increase in specific heat.  The remaining modes (that are not simulated) behave
essentially as an ideal gas.

As a physical example, consider our simulations for $\Cnl = 20000$. 
Choosing $L = 25 {\mu}$m, and Rb-87, this corresponds to approximately 
$3.8 \times 10^5$ atoms below the cutoff satisfying $N_k > 10$ at $T_c$ of about 370
nK.  There are about $1.3 \times 10^6$ atoms in total, with a total number
density of  $8.3 \times 10^{13}$ cm$^{-3}$. Thus in this situation
nearly 30\% of the atoms are simulated by the PGPE.

\subsection{Comparison of methods of temperature determination}

As described in Sec.~\ref{blergh} there are many choices of the operator $\D$
that may be used in Eq.~(\ref{eqn:T}).  The resulting calculations only
give the same temperature if the system is in equilibrium, so this provides an
important confirmation that the system has thermalized. In this paper we
consider two cases   $\D_Q= \{ \partial/\partial Q_i,\,0\}$ and $\D_P= \{
0,\partial/\partial P_i\}$, and we refer to the temperatures calculated from
these operators as $T_Q$ and $T_P$ respectively. Allowing $Q$ or $P$
derivatives only in the separate definitions of the operator $\D$ simplifies
the calculation of Eq.~(\ref{eqn:T}) due to the elimination of mixed
derivatives.

\begin{figure}
\includegraphics{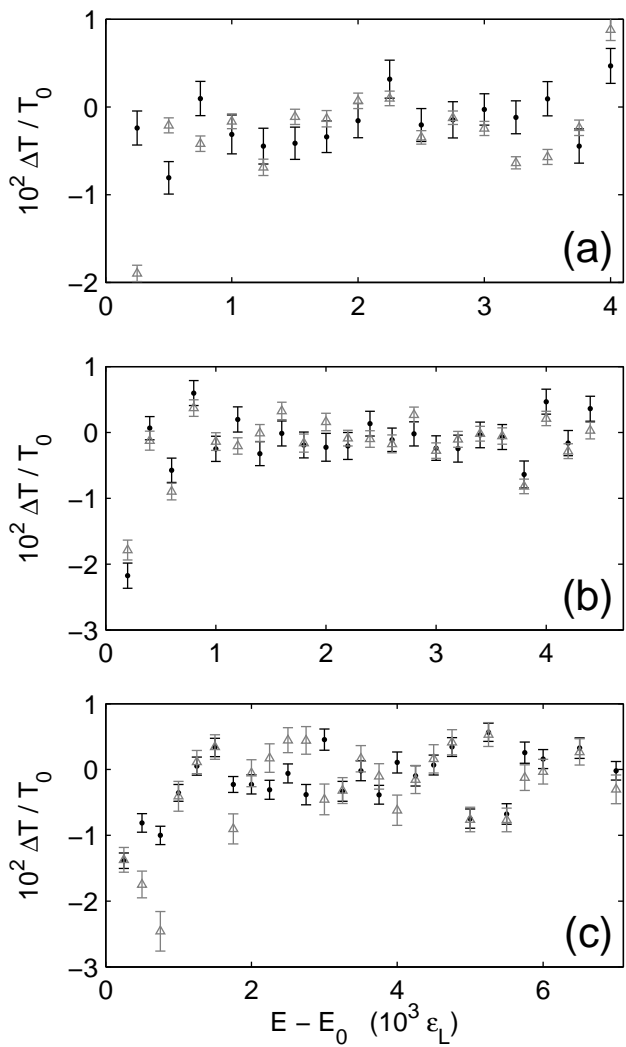}
\caption{Plot of the relative differences of simulation temperatures $T_Q$ and
 $T_P$ calculated from
Eq.~(\protect\ref{eqn:T}) with temperatures $T_0$ determined from the same data plotted in
Fig.~9 of Ref.~\protect\cite{davis2002}.  
Grey triangles: $\Delta T = T_P - T_0$, black dots: $\Delta T = T_Q - T_0$.  (a) $\Cnl$ = 500,
(b)  $\Cnl$ = 2000, (c) $\Cnl$ = 10000.}
\label{fig:compare}
\end{figure}

We begin by comparing $T_Q$ and $T_P$ with previous results from
Ref.~\cite{davis2002}. In this earlier work we obtained temperatures using three
different methods, two based on Bogoliubov quasiparticles and perturbation theory
and a third non-perturbative calculation. This third method did not have a firm
theoretical basis; however, we showed that the results were consistent with the two
other calculations in their regime of validity, and gave reasonable results more
generally. Figure~\ref{fig:compare} shows the relative differences between the new
simulation temperatures $T_Q$ and $T_P$ calculated from Eq.~(\protect\ref{eqn:T})
and the temperatures $T_0$ determined from the earlier method three. The simulation
data used is the same as that plotted in Fig.~9 of Ref.~\protect\cite{davis2002}.

We can see from Fig.~\ref{fig:compare} that only a small number of points
differ by more than one percent from the  previously determined values, and even
these would be hard to detect on a plot of the absolute temperatures. These results
therefore validate our earlier non-perturbative method for temperature determination
in a homogeneous system (described in Sec.~VI.D of Ref.~\cite{davis2002}).
Figure~\ref{fig:compare} also shows that in general the values of $T_Q$ and $T_P$
agree with each other within their error bars.  The error is determined from the
standard deviation of the expectation value of Eq.~(\ref{eqn:T}) divided by the
square root of the number of samples (in this case two hundred).  This estimate
assumes gaussian statistics, which seems reasonable when the distribution of values
is plotted as a histogram; however, it may underestimate the actual error somewhat. 
The agreement between these distinct determinations of temperature confirms their
validity and provides important further evidence that the PGPE evolves randomised
initial states to a thermodynamic equilibrium consistent with the microcanonical
ensemble.

\subsection{Shift of the transition temperature}
\label{sec:Tshift}

\begin{figure}
\includegraphics{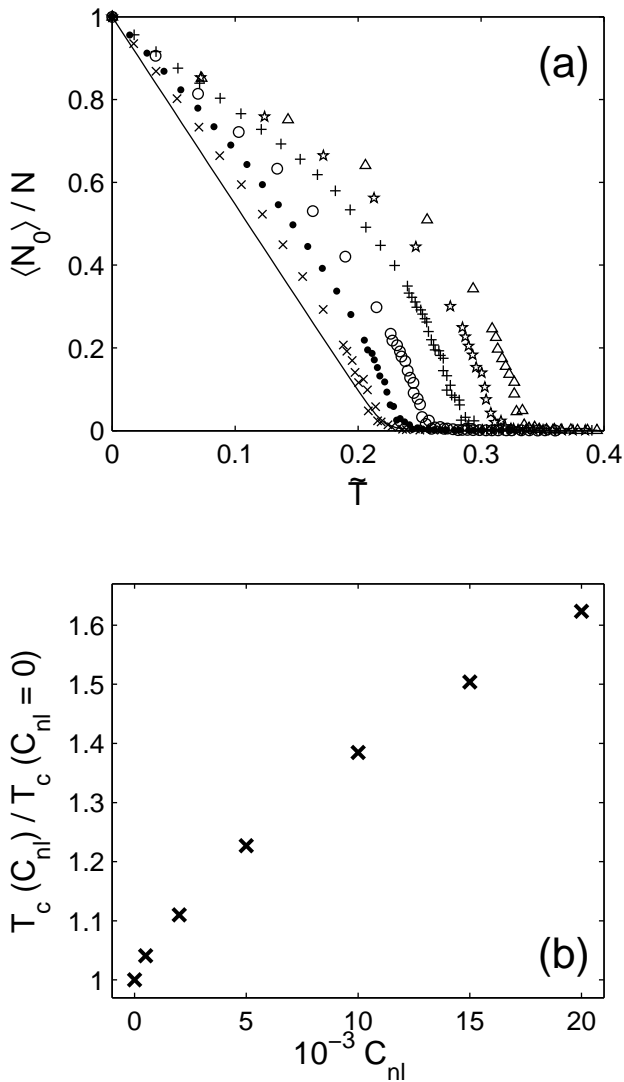}
\caption{(a) Plot of the condensate fraction versus temperature for a number of
interaction strengths.  Solid line $\Cnl = 0$, crosses $\Cnl = 500$,
solid dots $\Cnl = 2000$, open circles $\Cnl = 5000$,
plusses $\Cnl = 10000$, stars $\Cnl = 15000$, open triangles $\Cnl = 20000$.
(b) Plot of the transition temperature versus interaction strength.  The
transition temperature is determined by the method of Binder cumulants as
described in the text.
}
\label{fig:tshift}
\end{figure}

Figure \ref{fig:tshift}(a) plots the equilibrium temperatures and condensate
fractions for several series of simulations with different nonlinearities
$\Cnl$, as well as for the ideal gas.  These can be interpreted to be
simulations at a fixed density with a varying scattering length.  It is
immediately obvious that qualitatively the transition temperature increases
with increasing nonlinearity, and this was noted in \cite{davis2002}.  Much
more data has been collected for this paper, and we now have a much more
reliable measure of temperature.  Thus we can now look at the shift of the
critical temperature with the nonlinear parameter $\Cnl$ for our PGPE system.

We can calculate the transition temperature for a non-interacting gas with
equipartion occupation numbers and a
momentum cutoff $k_c$ in the continuum limit via
\begin{equation}
N = N_0 + \int_0^{k_c} \frac{d^3k}{(2\pi)^3} \frac{k_B T}{\hbar^2 k^2/(2m) - \mu}.
\label{eqn28}
\end{equation}
We find that the dimensionless critical temperature for a homogeneous 
PGPE system with a momentum cutoff of $k_c = 2\pi\kappa/L$ is 
$\tilde{T}_{c}(\Cnl=0) = \pi/\kappa$, where the dimensionless temperature is defined by
$\tilde{T} = k_B T / (N \epsilon_L)$.

Identifying the critical point in a finite-sized system with interactions,
however,  is somewhat more difficult.  Here we make use of the method of Binder
cumulants \cite{binder1981}, which have been used in other finite-size calculations
for the Bose gas \cite{Arnold2001b}. We note that the theory behind
Binder cumulants is derived entirely from canonical statistical  mechanics.  However,
the calculations of Caiani \etal.\  \cite{caiani2d,caiani3d} suggest that it is valid
as a numerical tool in the microcanonical ensemble, and we shall follow their
lead. The Binder cumulant can be
written as
\begin{equation}
C = \frac{\langle N_0^2 \rangle}{\langle N_0 \rangle^2},
\label{eqn:binder}
\end{equation}
where $N_0$ is the population of the zero momentum 
condensate mode in our simulations.
This quantity changes smoothly from one for the condensed system (ordered phase) to
two for the uncondensed system (disordered phase), with the width of the transition
region decreasing with increasing lattice size.  However, in lattice field theory
the chemical potential at which curves of $C$ vs $T$ intersect for different lattice sizes is
universal for a given universality class, which is three-dimensional XY for our
system.  It has been calculated by Campostrini \etal.\ \cite{Campostrini} that this critical value is
$C_c = 1.2430(1)(5)$, where the first number in parentheses is due to statistical
errors and the second is due to systematic errors.

We therefore determine the critical temperature from our simulations by finding the
energy at which the Binder cumulant takes the value $C_c$ in equilibrium.  Due to
our limited statistics from 200 field samples, the results are somewhat noisy, but
we are able to identify $\tilde{T}_c$ for the simulations to an accuracy of
approximately one percent.

We note that for the case of $\Cnl=20000$, the predicted shift in critical
temperature is more than 60\%.  However, this corresponds to the shift in
\emph{dimensionless} temperature of the low-energy states, not the shift in
the critical temperature of the complete system which will be smaller. 
This will be discussed in more detail in Sec.~\ref{sec:Tcshift}.

\subsection{Calculation of the specific heat}
Although the specific heat can theoretically be determined by a similar
procedure to that used for the temperature, the actual formulae are rather
complicated and difficult to calculate.  Instead in this section we use
numerical methods to calculate curves for the specific heat.

\begin{figure}
\includegraphics{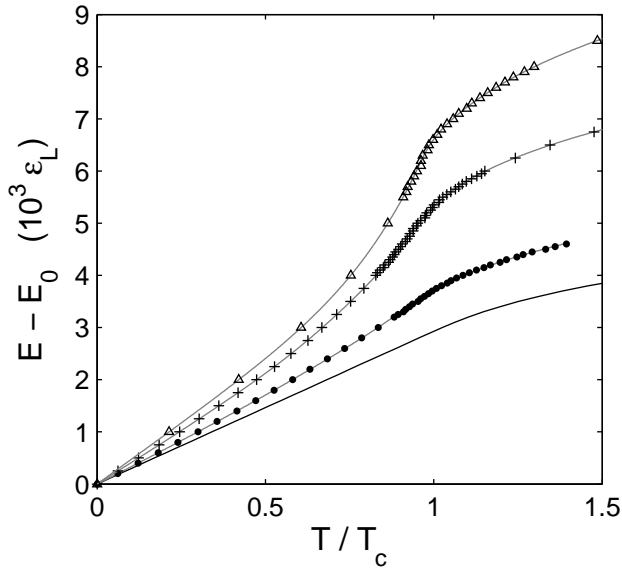}
\caption{The system energy plotted against the temperature, scaled in units of
the critical temperature $T_c$ so that the curves are distinct in the linear region.
Solid line $\Cnl = 0$, 
solid dots $\Cnl = 2000$, 
plusses $\Cnl = 10000$, open triangles $\Cnl = 20000$. The grey lines are the
smoothing spline fits to the numerical data.
}
\label{fig:EvT}
\end{figure}

The calculation of numerical derivatives is difficult for data with statistical
errors.  Here we have used a smoothing spline fitting technique to the raw
numerical data for energy and temperature, and calculated the derivative from
this fit.  Examples of the spline fits to the numerical data are plotted in
Fig.~\ref{fig:EvT}.

\begin{figure}
\includegraphics{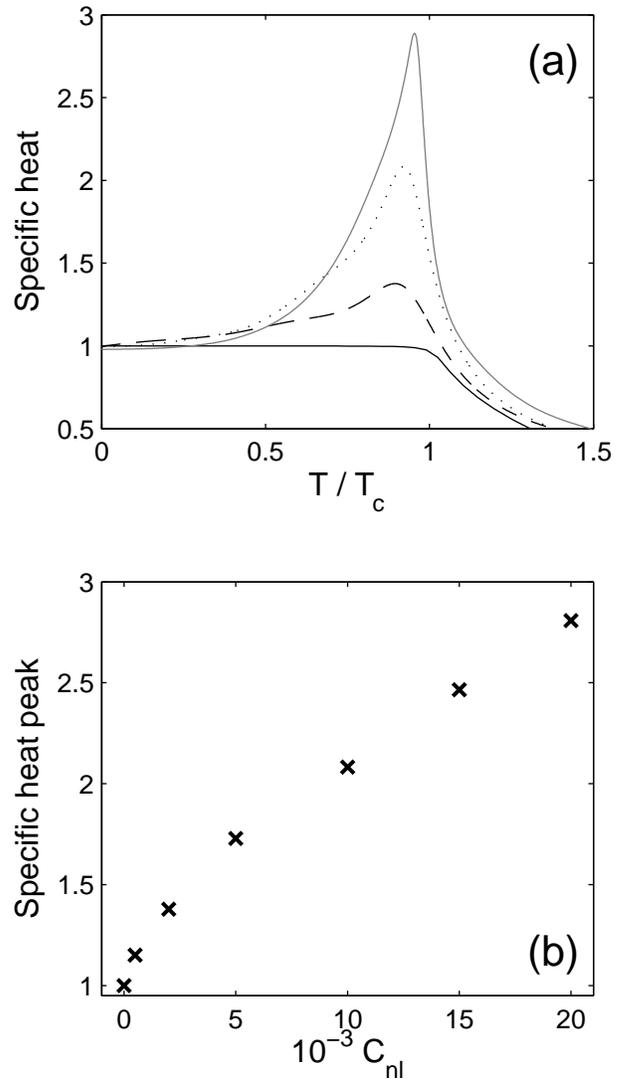}
\caption{(a) The numerically calculated specific heat curves for various interaction
strengths.  The peaks occur at temperatures a few percent below the identified
transition temperature.  We estimate the error for these curves to be of order 
a few percent.  Solid black line  $\Cnl = 0$, 
dashed line $\Cnl = 2000$, 
dotted line $\Cnl = 10000$, solid grey line $\Cnl = 20000$.
(b) The maximum value of the specific heat plotted versus
the dimensionless interaction strength $\Cnl$.  For both (a) and (b), the specific heat is plotted relative to the corresponding
value at $T=0$, and so the quantities are dimensionless.
}
\label{fig:spheat}
\end{figure}

The specific heat curves calculated from the data in Fig.~\ref{fig:EvT} are
shown in Fig.~\ref{fig:spheat}(a).  The units of the vertical axis are scaled
by the specific heat of the ideal Bose gas for the same system at $T=0$.  We
can see that there is a strong peak near the critical temperature that
increases with increasing $\Cnl$.  Scaling theory for critical points in the
thermodynamic limit suggests that the specific heat will be discontinuous at
the phase transition. In our case the peak is not exactly at $T_c$, as
perhaps would be expected. We presume that this is due to to a combination of
finite size effects and numerical errors in the fitting procedure, which we
estimate to be a few percent.  Similar behaviour has also been noted in
\cite{caiani3d}. Figure~\ref{fig:spheat}(b) shows the maximum value of the
specific heat plotted versus $\Cnl$.

\section{Relation to other work}
\label{sec:other}

\subsection{Dynamics of $\phi^4$ lattice field theory} 
\label{sec:dynphi4}

The results presented in this paper for the homogeneous Bose gas have many
similarities to classical $\phi^4$ lattice field theory, which is often studied in
relation to second order phase transitions.  In such studies the field is
discretised on a lattice with the spatial derivatives of the Hamiltonian being
approximated by finite difference methods.  Monte Carlo simulations are then
performed to study the thermodynamics.

However, there have also been ``molecular dynamic'' simulations of such
field theories, and in particular we note the work of Caiani \etal., who have
considered the phase transition via dynamical simulation of the $\phi^4$
model in both two \cite{caiani2d} and three dimensions \cite{caiani3d}.  Their
equations of motion are distinct from those of this paper by virtue of being  
second order in time.  Their paradigm Hamiltonian in $d$-dimensions is
\begin{equation}
H[\phi] = \int d^d\mathbf{x} \frac{1}{2} \pi^2(\mathbf{x}) + \frac{J}{2}[\nabla
\phi(\mathbf{x})]^2 + \frac{1}{2} \phi^2(\mathbf{x}) + \frac{\lambda}{4}
\phi^4(\mathbf{x}),
\end{equation}
with the canonical position variables $\phi(\mathbf{x})$ and conjugate momenta
$\pi(\mathbf{x}) = \dot{\phi}(\mathbf{x})$, where $\phi$ is a vector quantity with
up to four dimensions.   We note that as
 this Hamiltonian is of the form
  $H = \pi^2/2 + V(\phi)$,  both the temperature and specific heat of these 
simulations can be calculated from expectation values of the kinetic
energy.  This is not possible for the Hamiltonian we consider in this paper
where the interaction term mixes powers of the position and momentum
co-ordinates.

Also, in  Ref.~\cite{caiani2d} the parameters used were $J = 1, \lambda = 0.6$,
and for Ref.~\cite{caiani3d} the values $J = 1, \lambda = 0.1$ and $\lambda = 4
$ are specifically mentioned.  Thus these calculations appear to be in quite a
different regime to the results presented here. Despite these differences,
however, it seems that much of their numerical data is
qualitatively similar to ours.

\subsection{Shift of $T_c$ in the continuum limit}
\label{sec:Tcshift}

The results presented in this work can also be connected to the issue of the
shift in 
the transition temperature for the homogeneous Bose gas, which has been the
subject of a number of recent papers.  In the weak interaction limit the shift
$\Delta T_c$ has the form
\begin{equation}
\frac{\Delta T_c}{T_{c0}} = c a n^{1/3}.
\label{eqn:density}
\end{equation}
where $T_{c0}$ is the transition temperature for the ideal gas, $n$ is the
density, $a$ is the $s$-wave scattering length and $c$ is a dimensionless
constant.  The value of $c$ cannot be determined by
perturbation theory as this breaks down at second-order phase transitions due
to infrared divergences.    There have been several calculations of the value
of $c$, differing by up to an order of magnitude and even in sign (see the
summary in \cite{Arnold2001}).

The dimensionless constant $c$ has recently been determined via Monte Carlo calculations by
Arnold and Moore  \cite{Arnold2001,Arnold2001b} and Kashurnikov \etal.\
 \cite{Kashurnikov2001} to be $c = 1.32 \pm 0.02$ and $c = 1.29\pm 0.05$
respectively.  These calculations were carried out via classical $\phi^4$ field
theory, which can be systematically matched to the problem of the homogenous
interacting Bose gas.  The Monte Carlo calculations proceeded 
by sampling the classical action  
\begin{equation}
\frac{S}{\beta} =  \int d^3\mathbf{x} \left[ \psi^*(\mathbf{x}) \left(-\frac{\hbar^2\nabla^2}{2m} -
 \mu_{\rm eff}\right) \psi(\mathbf{x})
+ \frac{U_0}{2} |\psi(r)|^4\right],
\label{eqn:action}
\end{equation}
on a lattice at a fixed temperature $T$, where $\beta = (k_B T)^{-1}$.  
The value of  $\mu_{\rm eff}$ was adjusted until the
critical point was reached, and thus the shift in critical density 
$n_c = \langle |\psi|^2 \rangle$ from the ideal gas value $n_{c0}$
could be measured.  The shift in critical temperature at a fixed density can then be
determined from
\begin{equation}
\frac{\Delta T_c}{T_{c0}} = -\frac{2}{3} \frac{\Delta n_{c}}{n_{c0}},
\label{eqn:Tcshift}
\end{equation}
which is easily derived from the formula for the critical temperature of the ideal
gas.
While this procedure seems straightforward, in practice it is necessary to 
give careful consideration to 
finite-size effects in the calculation---see Ref.~\cite{Arnold2001b} for a
detailed discussion of these matters.

The results of simulations similar to those presented here  can also be used to
calculate a value for $c$, as we are also sampling the thermodynamic functions of
classical $\phi^4$ field theory.  The Monte Carlo calculations fix the temperature
and adjust the value of $\mu_{\rm eff}$ in  Eq.~(\ref{eqn:action}) which then
determines the normalisation of the field.  In our calculations, we adjust the
energy of the initial state to find the critical point and determine the temperature
using the method described above.  Our simulations have a fixed normalisation, but
the dimensionless temperature $\tilde{T}\propto T/N$, so for a given value of $\Cnl$
we can interpret our results as being at a fixed temperature and a varying density.

The main difference between the methods is the manner in which field
configurations are sampled.  The Monte Carlo methods can use 
the most efficient update possible, as long as the samples are canonical at
a given temperature.  Our calculations solve for the evolution of a
microcanonical field, and use the theorem of ergodicity to generate an
ensemble.   We have one minor advantage in that our momentum cutoff is
spherically symmetric, whereas the Monte Carlo calculations simulate the first
Brillouin zone of the lattice. However, the molecular dynamics method suffers
from critical slowing down---as the energy of the highest modes is $\propto
k^2$, we require time steps of order $\delta t = 1/k_c^2$ where $k_c$ is the
momentum cutoff.  Thus our simulations are disproportionally less efficient for
larger grids compared to the Monte Carlo calculations, and will not be able to
generate results as accurately for a given computation time 
\cite{Moore_private}. Nonetheless, we can use our simulations to confirm
qualitatively the results of the Monte Carlo analysis, providing an independent
demonstration of the validity and potential usefulness of our temperature
determination.

\begin{figure}
\includegraphics{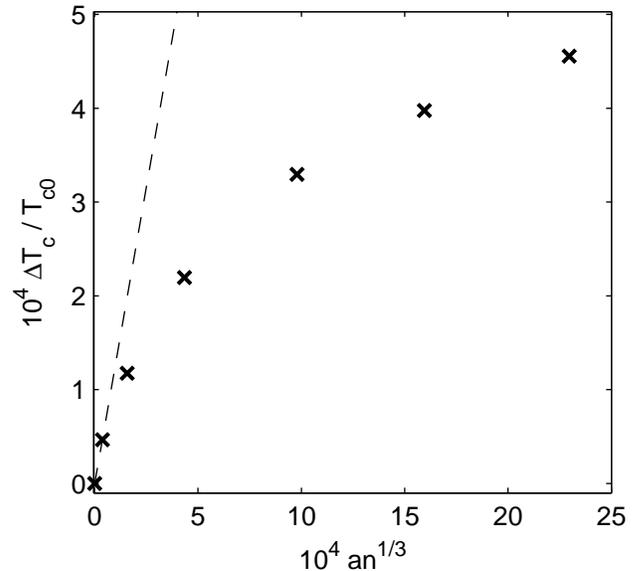}
\caption{Shift in the critical temperature with interaction strength determined from
the results presented in this paper with $N_{\rm below}(\Cnl=0) = 10^{10}$.  The dashed line
is a linear fit to the first two data
points and this has a slope of $ 1.3 \pm 0.4$.}
\label{fig:shift}
\end{figure}

As a simplified illustration, we follow through the logical procedure that would be
required to calculate a value for $c$.  To consider the shift in the critical
point, we can consider the shift in the critical density given a fixed critical
temperature $T_0$.   In our numerical simulations
we keep $\Cnl = 8 \pi a N_{\rm below}/L$ fixed and measure
a shifted critical temperature 
\begin{equation}
\tilde{T}_c = \frac{k_B T_0}{N_{\rm below} \epsilon_L}.
\end{equation}
Here $N_{\rm below}$ is the number of particles below the cutoff. If
we fix the critical temperature at $T_0$ as well as the system size $L$
(and hence $\epsilon_L$), we can interpret the increase in the dimensionless quantity
$\tilde{T}_c$ as a decrease in the value of $N_{\rm below}$ and hence a decrease in 
the critical density.  The most important point to note is that as long as we have
$k_c \gg k_0$, where $k_0$ labels the division between quasi-particle and particle
like excitations at the transition temperature, then particles above the cutoff will
not be significantly affected by the change in the interaction strength.

We therefore calculate $N_{\rm above} = N_{\rm tot} -
N_{\rm below}$ for the ideal gas, 
where $N_{\rm tot}$ is the total number of particles. This will be a constant as long as $k_c \gg
k_0$.  We can then calculate
$N_{\rm below}(\Cnl)$ and hence the shift in the critical density 
from the simulation data, and by using the relation of Eq.~(\ref{eqn:Tcshift}), 
we obtain
the shift in the critical temperature.

This can then be plotted against $an^{1/3}$ and the slope at the origin
determines the coefficient $c$.  This plot is given in Fig.~\ref{fig:shift},
where we have set $N_{\rm below}(\Cnl=0) = 10^{10}$. We note that the method
does not depend on the value chosen for $N_{\rm below}(\Cnl=0)$ as long as it
is large enough that $\langle N_k \rangle \gg 1$ is well satisfied.

By fitting a straight line to the first two points as illustrated in Fig.~\ref{fig:shift},
we get an estimate for the coefficient
\begin{equation}
c = 1.3 \pm 0.4,
\end{equation}
where the error specified is due to the uncertainty in the value of $T_c$ for the
data point. This agrees with the value determined in
Refs.~\cite{Arnold2001,Kashurnikov2001} --- a result that 
should be treated with caution.
The correct
value of $c$ will only be reached in the limit of large volume and small lattice
spacing, and we believe we have not reached this regime.
  For comparison with the results of
Arnold and Moore, for our first data point we have $L u \approx 325$ and  $ u a_{\rm
latt} \approx 10.2$, where $u = 3 \tilde{T} \Cnl / L$ and $a_{\rm latt} = L/32$. 
Our other data points have values for these quantities that are
much larger than this.  Arnold and Moore suggest
that $Lu \ge 400$ and $u a_{\rm latt} \le 6$ are necessary to get 
an accurate result for $c$ without a finite-size scaling analysis 
\cite{Arnold2001b,Moore_private}.

We could potentially improve our results by performing such a
finite-sized scaling analysis, but there is little reason to do so given the
greater accuracy obtained in Refs.~\cite{Arnold2001,Kashurnikov2001}.  The purpose
of this calculation is to demonstrate a useful application of our  temperature
determination with the PGPE in a non-perturbative regime.  In this regard the
qualitative agreement with earlier more involved and specialized calculations
provides a pleasing confirmation of the general validity of the method.


\section{Conclusions}
\label{sec:conc}

We have shown that the projected Gross-Pitaevskii equation can be exactly
mapped to Hamilton's equations of motion for canonically conjugate position and
momentum variables.  Using this mapping we have described how to utilise the
microcanonical thermodynamic method of Rugh \cite{rugh01} to measure the
temperature of PGPE simulations in the non-perturbative regime.  This method
agrees with previous calculations described in Ref.~\cite{davis2002}, but has a
rigorous theoretical justification and wider applicability.   Using this
approach, we have quantitatively measured the shift in the critical temperature
for condensation with the nonlinear constant $\Cnl$.  We have also observed
that the specific heat reaches a maximum near the transition point as expected
from the theory of continuous  phase transitions, and that the peak value
increases with the nonlinearity.  Finally, we have made a connection between
these calculations and Monte Carlo simulations that have determined the shift
in the critical temperature with scattering length of the homogeneous Bose gas
in the continuum limit.  This is further evidence that the projected GPE should
be valid for dynamical calculations through the critical region as long as the
condition on the occupation numbers is satisfied.

\begin{acknowledgments} 

The authors are grateful to Peter Arnold and Guy Moore for their insightful
comments on the relation of this work to that of
Refs.~\cite{Arnold2001,Arnold2001b}.  We also thank Crispin Gardiner for his
useful comments.  MJD would like to thank Tim Vaughan, Karen Kheruntsyan, Joel
Corney, and Peter Drummond for several useful discussions at various stages of
this work.  MJD acknowledges the financial support of the University of
Queensland and the Australian Research Council Centre of Excellence grant
CE0348178.  SM would like to thank the Royal Society of London for
financial support.

\end{acknowledgments}

\end{document}